\documentstyle[prl,aps,epsf]{revtex} 
\input epsf

\def\({\left(}  
\def\){\right)}

\begin{document}
\draft
\twocolumn[\hsize\textwidth\columnwidth\hsize\csname@twocolumnfalse%
\endcsname

\preprint{}
\title{Quasiparticles in the multicomponenet Zhang-Hansson-Kivelson model}
\author{  D.~Serban}
\address{
{\it Service de Physique Th\'eorique, 
CE-Saclay, F-91191 Gif-sur-Yvette, France.}
}

\date{\today}
\maketitle

\begin{abstract}


We study the vortex solutions in a multicomponent 
Zhang-Hansson-Kivelson model for the fractional quantum Hall effect,
at the self-dual point. Vortices
with minimal free energy represent Laughlin quasiholes.
We find at least two classes of solutions, distinguished by 
their global invariance, or by the number of conserved charges.

\end{abstract}
\vspace{0.5cm}
]

\section{Introduction}

Studying properties of the quantum Hall systems requires to know the properties
of the elementary excitations. There exist several approaches to the
question of quasiparticles in quantum Hall effect. One has as starting 
point the Laughlin wave function \cite{bob}. On the other hand, low
energy excitations located at the edges can be described by means
of conformal field theory \cite{wen}. 
As for the excitations in the bulk, they can be obtained in field theories
with Chern-Simons term. A particularly useful effective theory for the bulk, 
inspired by the Landau-Ginsburg theory of the 
superconductivity, is the so-called Zhang-Hansson-Kivelson (ZHK) model \cite{zhk}. 
It consists of a complex scalar field coupled to a statistical gauge
field, which performs the statistical transmutation from bosons to fermions.
Similarly to the Landau-Ginsburg model for the superconductivity \cite{sarma}, the ZHK
model shows vortex solutions, vortices being 
identified to the Laughlin
quasiparticles (or, more precisely, to the Laughlin quasi-holes, since they
correspond to a depletion of the electron density). 
The vortices carry (fractional) electric charge, 
as well as statistical flux.
A pedagogical introduction to Chern-Simons theories, as well
as to vortex solutions in field theories with Chern-Simons 
term can be found in ref. \cite{dunne}.

In this letter, we study the vortex solutions in the situation 
where the bosonic field has 
$p$ components. 
The relevant filling fractions are of the type $\nu=p/(p\beta+1)$, 
where $\beta=2n$ is an even integer (the number of fluxes attached to
the electrons in the Jain construction).
Let us point out that these filling fractions do not completely characterize the
quantum Hall fluid; for example, $p=2s$ even integer
can correspond to the spin polarised fractional quantum Hall effect,
{\it or} or to the spin singlet fractional quantum Hall effect \cite{lopfr}.

Pursuing the analogy with the Landau-Ginsburg theory for the
superconductors \cite{sarma,dunne}, we minimize the free energy at the self-dual
point, where vortices do not interact. This point has also the 
merit that the minimum of the free energy is solution of  
first order differential equations \cite{bogo}.
For superconductors, the dual point correspond to the border
between type I and type II superconductors \cite{sarma}, and above 
the dual point vortices interact repulsively.

When $p>1$, we find that the vortex equations have several 
types of solutions, depending
on the monodromy conditions around the vortex center
imposed to the electron fields. Different monodromy conditions
lead to different number of independent electron
densities, which ultimately lead to different number of conserved charges.
Therefore, the bulk action of the ZHK model does not uniquely
characterize a quantum Hall fluid, and 
different types of solutions
correspond to different quantum Hall fluids associated to
the same filling fraction.

For $p=2$, a similar study was done by Ichinose and Sekiguchi \cite{ichi},
who found two type of excitations: the ones which carry $su(2)$ charge
were called merons, while the singlet superposition of merons
was called vortex. Our first type of solutions correspond to the 
merons. The second type is similar to the vortex, in the sense 
that it carries only a $u(1)$ charge; however, it carries only a fraction
of the charge and the flux of the vortices in \cite{ichi}. 
As pointed out in \cite{ichi}, turning on an interlayer tunneling
(which break the symmetry from $su(p)$ to $u(1)$) renders the
meron excitations unstable. We therefore expect that the vortices
carrying fractional flux are good excitations for this kind of system.

\section{ZHK model with $p$ components}

Throughout this paper we  are going to use units in which $c=\hbar=1$.
Also, for simplicity we set $e=2m=1$.
The ZHK model \cite{zhk} couples a complex scalar field to a statistical
gauge field with Chern-Simons dynamics and it was originally proposed
to describe filling fraction of the type $\nu=1/(2n+1)$. 
More complicated filling fractions  
can be described by a complex scalar field with 
$p$ components, $\psi_k$, coupled to $p$ statistical gauge fields.
The action of the multicomponent ZHK model is then:
\begin{eqnarray}
\label{zhk}
{\cal S}=\sum_{k=0}^{p-1} \int d^3 x \bigg(- |\vec D^{(k)} 
	\psi_k(z)|^2 
	&+&i\psi^*_k(x)  D_0^{(k)} \psi_k(x)\bigg) \\ \nonumber
	-V[{\bf \psi}]&+&{\cal S}_{CS}[{\bf a}]\;,
\end{eqnarray}
where the covariant derivatives $D_\mu^{(k)}$ involve the 
external electromagnetic 
potential $A_\mu$
and the statistical gauge potentials $a_\mu^{(k)}$ :
	$$D_\mu^{(k)}=\partial_\mu +i  (A_\mu +  a_\mu^{(k)})\;,$$
$V[{\bf \psi}]$ is a two body potential whose form will be precised later, and 
${\cal S}_{CS}[{\bf a}]$ is a Chern-Simons term for the statistical 
gauge potentials.
The Chern-Simons term is supposed to be of the form \cite{wenzee}
	$${\cal S}_{CS}[{\bf a}]
	=-\frac{1}{4\pi} \int d^3x\; \epsilon^{\mu \nu \lambda} a_\mu^{(k)}
	(K^{-1})_{kl}\;\partial_\nu a_\lambda^{(l)}\;,
	$$
with $\epsilon^{\mu \nu \lambda}$ being the Levi-Civita tensor and
$K$ a $p\times p$ symmetric matrix with integer entries. We are interested
here in the case $K_{kl}=\beta +\delta_{kl}$, when the filling factor is
$\nu=\sum_{k,l}(K^{-1})_{kl}=p/(p\beta+1)$.

The time components of the gauge potentials play the role of Lagrange multipliers,
insuring the attachement of the statistical flux to the charge
	\begin{equation}
	\label{attach}
	\rho_k \equiv |\psi_k|^2 =-\frac{1}{2\pi} (K^{-1})_{kl}\; \epsilon^{ij}
	\partial_i a_j^{(l)}=-\frac{1}{2\pi} (K^{-1})_{kl} \;b^{(l)}\;.
	\end{equation}
Let us remind that the total (electric) charge density is $\rho=\sum_k \rho_k$.
	
This equation is solved in the gauge $a_0^{(k)}=0$ by
	$$a_i^{(k)}(x)=-K_{kl}\int d^2 \vec y\; \epsilon^{ij}\; 
	\partial_j \theta(\vec x-\vec y)\;
	\rho(y)\;,$$
where $\theta(\vec x)=\arctan(x_2/x_1)$ is the angle of the vector $\vec x$ 
in the plane.
For a collection of $N_k$ point particles of density $\rho_k(\vec x,t)=
\sum_{a=1}^{N_k} \delta (\vec x -\vec x_a^{(k)}(t))$, the gauge potentials 
$a_i^{(k)}$
could be removed via a singular gauge transformation. Under
this transformation, the wave function for the particles acquire
a phase factor 
	$$\varphi^{(k)}(\vec x)=\sum_{a,l} K_{kl} \;\theta(\vec x-\vec x_a^{(l)})
	\;.$$
Since the diagonal entries of the matrix $K$ are odd, this singular 
gauge transformation turns bosons into fermions.

\section{Static self-dual solutions}
\label{sol}

We consider a potential $V[\phi]$  which has 
a minimum at $\rho_k=n/p$, $n$ being the total electron density
	$$V[\psi]=\int d^2x \;d^2 y \;\delta \rho_k(x) \;
	\delta\rho_l(y)\;V_{kl}(x-y)\;,
	 $$
where $\delta \rho_k(x)=\rho_k(x)-n/p$ is the deviation of the
partial density from its average value.
For simplicity we use a hard core potential.
A self-dual point exists when
	$$ V_{kl}(x-y)={2\pi\;K_{kl}}\;\delta(x-y)\;.
	$$
For a constant external magnetic field, $\epsilon_{ij}\partial_iA_j=B_{ext}$
and $A_0=0$, 
the static free energy functional is
	\begin{eqnarray}
	\label{free}
	{\cal F}= \int d^2x \bigg(\sum_k \big| \big(
	\partial_i +i(A_i&+&a_i^{(k)})\big)\psi_k \big|^2\\ \nonumber
	&+& 2\pi \; K_{kl}\;\delta \rho_k\;
        \delta \rho_l\bigg)
        \end{eqnarray}
whose minimum is realized by the uniform configuration
$a_i^{(k)}=-A_i$, $a_0^{(k)}=0$, $\psi_k=\sqrt{n/p}$.
This solution is consistent with the attachement of the statistical flux
to the charge (\ref{attach}) if
	$$n= \sum_k\rho_k=\frac{\nu}{2\pi} B_{ext}\;,$$
that is, if the filling factor defined as the occupation of the
Landau level equals exactly $\nu=p/(p\beta+1)$.
Small deviations from this filling fractions can be accomodated
by creating locally non-uniform configurations.	
Let us now search such non-uniform configurations with finite energy. 
We transform the expression (\ref{free}) using the Bogomol'nyi identity
	$$|\vec D \psi|^2=|D_\pm \psi|^2\mp B |\psi|^2 \pm\epsilon_{ij}
	\partial_i J_j\;,$$
where $D_\pm=D_1\pm iD_2$ and $J_j=\big(\psi^*D_j\psi-\psi(D_j \psi)^*\big)/2i$.
After some algebra, and dropping the boundary contribution, we obtain that 
	$${\cal F}= \int d^2x \sum_k\bigg( |D_-^{(k)}\psi_k|^2+\
	\frac{\nu}{2\pi p}   B_{ext} B^{(k)})\bigg)\;,$$
where $B^{(k)}=B_{ext}+b^{(k)}$ is the total magnetic field (external and statistical) 
seen by the component $k$. Therefore, the free energy is bounded by a multiple of
the total magnetic flux. This bound is saturated by configurations with
	\begin{eqnarray}
	\label{bogo}
	D_-^{(k)}\psi_k&=&0\;,\\
	B^{(k)}=B_{ext}&-&2\pi\; K_{kl}\;\rho_l\;.\nonumber
	\end{eqnarray}
To solve these equations, we parametrize $\psi_k$ by the modulus and the phase
	$$\psi_k=\rho^{1/2}_k e^{-i\chi_k}\;.$$
Then, the first equation in (\ref{bogo}) become
	$$A_i+a_i^{(k)}=\partial_i \chi_k +\frac{1}{2} \epsilon_{ij} \partial_j 
	\ln \rho_k\:,$$
The two equations (\ref{bogo}) can be combined into a system
of $p$ coupled equations for the partial densities $\rho_k$
	$$\nabla^2 \ln \rho_k =4\pi\; K_{kl}\;(\rho_l-n/p)\;.$$
Localized (finite energy) solutions have $\rho_k \to n/p$ when $|x|\to \infty$,
as well as critical points $x_a$, where $\rho_k(x_a)=0$ for some $k$. Let us consider 
a
radially symmetric configuration with a critical point at $x_0=0$, and switch
to polar coordinates $(r,\theta)$.
This configuration is characterized by the vorticities $s_k$
	\begin{eqnarray*}
	\chi_k(\theta+2\pi)-\chi_k(\theta)=2\pi s_k\quad {\rm and} \\
	\rho_k^{1/2} (r ) \sim\;r^{s_k}\quad {\rm at }\quad r\to 0\;.
	\end{eqnarray*}
The magnetic flux carried by such a configuration is $\phi_k=2\pi s_k$ 
(the flux quantum is equal to $2\pi$).
The numbers $s_k$ are constrained via the monodromy conditions
around the vortex center imposed to the fields $\psi_k$.
When there is only one component, $p=1$, the phase can only have a jump of $2\pi \times
{\rm integer}$ around the center of the vortex, so that the field $\psi$
is single valued. When the electron field has several components, more complicated
monodromy conditions can appear.
The two typical situations which can arise are the following:
\begin{itemize}
\item{
$i)$ each component is single valued around the vortex center
	$$\psi_k(\theta+2\pi)=\psi_k(\theta)\;.$$
The minimal configuration is then of the type
$s_l=1$ and $s_{k}=0$ for $k\neq l$, so that the $l^{th}$ component 
carries one flux quantum and the others carry no flux. Vortex charges are 
given by
$q_k=K^{-1}_{kl}=\beta/(p\beta+1)-\delta_{kl}$.}
\item{
$ii)$ the phase is matched between two components after
a tour around the vortex center
	$$\psi_k(\theta+2\pi)=\psi_{k+1\;{\rm mod}\;p}(\theta)\;.$$
Then, there is only one independent density, $\rho_k=\rho_0$,
and it satisfies the 
differential equation 
	$$ \nabla^2 \ln \rho_0=4\pi (p\beta+1) (\rho_0-n/p)\;.$$
The vortex carries charge 
$(q_k=)q_0=-1/p(p\beta+1)$.
All $p$ components see the same flux, corresponding to
$s_0=\ldots=s_{p-1}=1/p$ and
the center of the vortex is a branch point singularity of order $p$.}
\end{itemize} 

Both above configurations have free energy ${\cal F}=n/2\pi p$.
In both cases, removing an electron of charge $1$ is equivalent to creating
$p\beta+1$ quasi-holes, and it requires free energy ${\cal F}=n/2\pi \nu$.

\section{Charge/neutral modes factorization}

In the definition of the ZHK model (\ref{zhk}), we have used the potentials $a_\mu^{(k)}$ .
They couple in a simple manner to the fields $\psi_k$, but are mutually coupled 
by the Chern-Simons part. Alternatively, we can choose to put the Chern-Simons
action into a diagonal form, by defining
	$$\alpha_\mu^{(k)}=\frac{1}{p} \sum_{l=0}^{p-1} e^{2i\pi kl/p} \;
	a_\mu^{(l)}\;.$$
Since $\alpha_\mu^{*(k)} =\alpha_\mu^{(p-k)}$, we obtain
	$${\cal S}_{CS}=-\frac{\epsilon^{\mu \nu \lambda}}{4\pi} \int d^3x\; 
	\bigg(
	\nu \;\alpha_\mu^{*(0)}\partial_\nu \alpha_\lambda^{(0)}
	+p \sum_{k=1}^{p-1} \alpha_\mu^{*(k)}\partial_\nu \alpha_\lambda^{(k)}
	\bigg)\;,$$
Also, we redefine densities
	$$\tilde\rho_k=\sum_{l=0}^{p-1}e^{2i\pi kl/p} \;
	\rho_l\;.$$
so that the flux to charge attachment is expresses in a diagonal form
	$$\tilde \rho_k=-\frac{c_k}{2\pi}\; \tilde b^{(k)}\quad {\rm with}\quad 
	c_k=\bigg\{ 
	\matrix{\nu\;, \quad k=0\cr p\;,\quad k\neq 0}$$
and $\tilde b^{(k)}=\epsilon_{ij}\partial_i \alpha_j^{(k)}$. 
This decoupling reveals the special role played by the mode, 
$\alpha_\mu^{(0)}=1/p\sum_k \alpha_\mu^{(k)}$, which will be called charge mode.
The other modes $\alpha_\mu^{(k)}$, $k=1,\ldots,p-1$ will be called
neutral. 
This charge/neutral mode separation is very similar to the
one employed by Balatsky and Fradkin \cite{bafr} in the context of
the singlet spin quantum Hall effect -- with the exception that here the 
neutral modes are described by an {\it abelian} Chern-Simons theory.

In this new basis, the charges carried by the two types of 
vortices are

\begin{itemize}
\item{
$i)$ electric charge $\tilde q_0=-1/(p\beta+1)$ and 
$\tilde q_k=e^{2i\pi kl /p}$ for the neutral modes $k\neq 0$,}
\item{
$ii)$ electric charge $\tilde q_0=-1/(p\beta+1)$ and 
$\tilde q_k=0$ for the neutral modes $k\neq 0$. }
\end{itemize}

The two types of vortices carry the same electric charge $\tilde q_0$,
but they have different neutral sectors: the first type carry a "spin" index,
while the second does not carry any other charge except the electric one.
Also, in this basis, the second type of vortex carries
only one type of flux, with value $1/p$ of flux quantum, which explains the branch
point singularity introduced at the center of the vortex.

An alternative point of view is that of the global invariance of $\psi_k$.
In the first case, there is a $u(1)^{\otimes p}$ global invariance,
corresponding to independent shifts of the phases $\chi_k$ (in fact, the 
invariance is extended to $u(1)\otimes su(p)$ \cite{froh}). In the second
case, the $p$ phases are not independent, and the global invariance is reduced 
to $u(1)$.
Intermediate cases are possible; for example, for $p=2s$ components,
the global symmetry can be arranged to be $u(1)\otimes su(2)$, 
as is the case in the spin singlet fractional quantum Hall effect \cite{lopfr}.

\section{Conclusion}

We conclude that the two different types of vortices described in section
\ref{sol} correspond to
two different types of quantum Hall fluids, both with the same value of the 
Hall conductivity but with different global invariance. 

The conformal field theory on the edge should have the same conserved charges as
the theory in the bulk. Therefore, we expect that the first class of solutions
correspond to an edge theory with the full $su(p)$ symmetry \cite{froh}.
In the second case, the global $su(p)$ symmetry should be broken, and
it has been shown \cite{twists} that one way this can be realized is by 
introducing twist operators. They are able to remove completely
the extra charges of the electron operators, as well as
the contribution of the neutral modes from the correlation functions.
The two types of conformal field theories, with and without the neutral modes, 
predict different 
behavior for the electron Green function \cite{kane,zuma,lopfr}, testable via 
tunneling conductance experiments \cite{gray}.

{\bf Acknowledgement} We thank V. Pasquier and K. Mallick for many useful 
discussions, and I. Ichinose for informing us about the reference
\cite{ichi} .   
 This work was partly supported by the TMR network contract
FMRX-CT96-0012 of the European Community.


\begin{references}

\bibitem{bob} R. Laughlin, Phys. Rev. Lett. {\bf 50}, 1395 (1983);
\bibitem{wen} X.G. Wen, Phys. Rev. B {\bf 41}, 12838 (1990);
 
\bibitem{zhk} S.C. Zhang, T. Hansson and S. Kivelson, Phys. Rev. Lett. 
{\bf 62}, 82 (1989); N.Read, Phys. Rev. Lett. {\bf 62}, 
86 (1989);
\bibitem{sarma} D. Saint-James, G. Sarma and E.J. Thomas, Type II superconductivity,
Pergamon Press 1969;
\bibitem{dunne} G. Dunne, {\it Les Houches lectures, 1998}, hep-th/9902115;
\bibitem{bogo} E. Bogomol'nyi, Sov. J. Nucl. Phys. {\bf 24}, 449 (1976);
\bibitem{wenzee} X.G. Wen and A. Zee, Phys. Rev. B {\bf 44}, 274   
(1991);
\bibitem{lopfr} A. Lopez and E. Fradkin, Phys. Rev. B {\bf 51}, 4347 (1995);
cond-mat/0008219;
\bibitem{ichi} I. Ichinose, A. Sekiguchi, Nucl. Phys B {\bf 493} (1997) 683;
\bibitem{froh} J. Fr\"ohlich and A. Zee, Nucl.Phys. B {\bf 364}, 517 (1991);
\bibitem{bafr} A. Balatsky and E. Fradkin, Phys. Rev. B {\bf 43}, 10622 (1991);
\bibitem{twists} V. Pasquier and D. Serban cond-mat/9912218;
G. Cristofano, G. Maiella, V. Marotta, 
cond-mat/9912287;
\bibitem{kane} C.N. Kane and M.P.A. Fisher, Phys. Rev. B {\bf 51}, 
13449 (1995);
\bibitem{zuma} A.V. Shytov, L.S. Levitov and B.I. Halperin,  
Phys. Rev. Lett. {\bf 80}, 141 (1998);  
U. Z\"ulicke and A.H. MacDonald, Phys. Rev. B {\bf 60}, 1837 
(1999);
D.H. Lee and X.G. Wen, cond-mat/9809160;
E. Fradkin and A. Lopez, Phys. Rev. B {\bf 59},
15323 (1999);
\bibitem{gray} A.M. Chang, L.N. Pfeiffer and K.W. West, 
Phys. Rev. Lett. {\bf 77}, 2538 (1996);
M. Grayson, D.C. Tsui, L.N. Pfeiffer, K.W. West
and A.M. Chang,  Phys. Rev. Lett. {\bf 80}, 1062, (1998);
  
\end{references}
\end{document}